\documentclass{article}

\usepackage{arxiv}

\usepackage[utf8]{inputenc} 
\usepackage[T1]{fontenc}    
\usepackage{hyperref}       
\usepackage{url}            
\usepackage{booktabs}       
\usepackage{amsfonts}       
\usepackage{nicefrac}       
\usepackage{microtype}      
\usepackage{lipsum}
\usepackage{graphicx}
\graphicspath{ {./images/} }

\usepackage{algorithmicx}
\usepackage{textcomp}
\usepackage{xcolor}
\usepackage{comment}
\usepackage{cleveref}
\usepackage[ruled, vlined, linesnumbered, norelsize]{algorithm2e}
\usepackage{algpseudocode}
\usepackage{pgfplots}
\usepackage{pgfplotstable}
\usepackage{caption}
\usepackage{subcaption}
\usepackage{todonotes}
\pgfplotsset{compat=1.17}
\usepackage{acronym}
\usepackage{enumitem}

\title{Decentralized Storage and Self-Sovereign Identity for Document-Based Claims}

\author{
 Bruno Gomes \\
  INESC-ID, Instituto Superior T\'{e}cnico,\\ Universidade de Lisboa \\
  \texttt{bruno.s.gomes@inesc-id.pt} \\
   \And
 Samih Eisa \\
  INESC-ID, Instituto Superior T\'{e}cnico,\\ Universidade de Lisboa  \\
  \texttt{samih.eisa@inesc-id.pt} \\
   \And
 David R. Matos \\
  INESC-ID, Instituto Superior T\'{e}cnico,\\ Universidade de Lisboa \\
  \texttt{david.r.matos@tecnico.ulisboa.pt} \\
  \And
 Miguel L. Pardal \\
  INESC-ID, Instituto Superior T\'{e}cnico,\\ Universidade de Lisboa \\
  \texttt{miguel.pardal@tecnico.ulisboa.pt} \\
}

\begin{document}

\maketitle
\begin{abstract}
Users increasingly rely on identity providers for accessing online services and resources. However, centralized identity systems often compromise user privacy due to online activity tracking or data breaches. At the same time, may online services require digital copies of physical documents for validation in claims processes, such as providing proof of residence for opening a bank account or verifying medical images for health insurance claims. With centralized solutions, privacy depends entirely on the trusted party, but there are emerging decentralized approaches that offer greater transparency. 

This article introduces \textit{SoverClaim}, a decentralized application prototype that empowers users to control their identity and also allows them to present digital documents with privacy. SoverClaim leverages \textit{Hyperledger Indy}, a blockchain for issuing and presenting self-sovereign digital identities with transparent audit logs, and \textit{Storj}, a decentralized peer-to-peer service, for secure and decentralized document storage and subsequent deletion. The prototype demonstrates the seamless integration of self-sovereign identities and document-based claims, achieving response times of under 750~ms, making it suitable for timely human interactions.
\end{abstract}

\keywords{Decentralized Application \and Peer-to-Peer \and Storj \and Self-Sovereign Identity \and Blockchain \and Hyperledger Indy}

\section{Introduction}
\label{sec:intro}

In contemporary digital systems, the reliance on identity providers to access online services and resources has become ubiquitous. These identity providers, which authenticate and authorize users, serve as important intermediaries between individuals and the many online platforms they wish to access. Despite their widespread adoption and the convenience they offer, centralized identity systems pose significant privacy concerns~\cite{twitter_news}. These concerns primarily stem from the potential for online activity tracking and the risk of data breaches, both of which can lead to unauthorized access to sensitive user information~\cite{data_leak_news}.

The issue of privacy is further complicated by the frequent requirement for users to submit digital copies of physical documents during various verification processes~\cite{ssi_evaluation_comparison}. For instance, when opening a bank account, individuals may be required to provide proof of residence, such as a utility bill or rental agreement. Similarly, health insurance claims often necessitate the submission of medical images or other sensitive health documents for validation. These processes, while necessary for identity verification and claims adjudication, expose users to additional privacy risks, particularly when centralized identity providers are involved.

In response to these challenges, decentralized approaches to identity management have begun to emerge. 
In particular, Self-Sovereign Identity (SSI)~\cite{Mhle2018ASO} emphasizes individual control over personal data, unlike traditional systems that rely on centralized authorities. 
SSI requires a robust infrastructure for the storage, management, and verification of digital identities and credentials.
Blockchain technology provides a suitable framework for managing SSI identities~\cite{ssi_survey}, supporting decentralized identifiers (DIDs) and verifiable credentials (VCs) registered, issued, verified, and revoked through smart contracts (code executed in the blockchain), without a central authority.
Blockchain data is immutable and every transaction can be transparently audited because data on it cannot be tampered.

Regarding document presentation, documents must be stored for the duration of the verification process and deleted afterwards to preserve privacy and comply with regulations such as GDPR~\cite{gdpr}, which mandates the ``right to be forgotten". 
However, blockchain is not suitable for storing these documents because of immutability. 
This is where decentralized storage systems, also known as peer-to-peer storage, become relevant~\cite{daniel2022ipfs}. 
These services fit well in the SSI base architecture by eliminating reliance on centralized storage, distributing the documents for several nodes, which enhances data resilience. However, not all decentralized storage solutions provide deletion assurances.

By integrating blockchain and decentralized storage technologies, we can envision a decentralized application (dApp) platform that prioritizes user control, privacy, and security while providing essential services. Nevertheless, specific implementations are required to realize these applications and assess their potential advantages.

This paper proposes SoverClaim, a prototype decentralized application, to demonstrate authentication with SSI and support secure document storage with sharing and deletion capabilities.
The name is meant to evoke ``sovereign claims'', in the sense that the user will keep control of his identity and claim documents. 
SoverClaim differentiates itself in four ways. First, it consolidates key functionalities of existing systems, such as creation, presentation, and encryption of digital identities. 
Second, it enables efficient and secure storage of identity documents with sharing and deletion.
Third, it further ensures accountability and auditability through decentralized identifiers and audit logs.
Finally, it adopts standard protocols for the creation of identification, with a strong emphasis on minimizing user data disclosure when accessing online services.

The remainder of this paper is structured as follows. Section~\ref{sec:bg-rw} discusses existing SSI management systems and decentralized storage, as well as existing works; Section~\ref{sec:soverclaim-overview} presents an overview of SoverClaim, followed by Section~\ref{sec:impl} that details the technical design and implementation. Section~\ref{sec:eval} presents the experimental results.

\section{Background and Related Work}
\label{sec:bg-rw}

This Section provides background details on SSI, blockchain technology and on decentralized storage systems.

\subsection{Self-Sovereign Identity}
\label{sec:ssi}

Self-Sovereign Identity (SSI) is a paradigm shift in identity management by placing control of personal data in the hands of individuals.
This approach challenges traditional centralized identity systems by empowering users to manage their digital identities independently~\cite{Buincu2024DesignAO}.
The core concepts of SSI include Digital Identity, DID, VC, VP, and Wallet.

A \textit{Digital Identity} is a representation of an individual or entity in the digital world, comprised of attributes such as name, date of birth, address, and other relevant information. Traditional digital identities are often managed by centralized authorities, such as governments or corporations.

\textit{Decentralized Identifier (DID)}~\cite{w3c_did} is a globally unique identifier designed to allow individuals and organizations to create their identifiers without a third party involved.
These identifiers enable users to demonstrate their ownership by utilizing cryptographic functions, which verify their possession of the corresponding private key associated with the DID.

\textit{Verifiable Credential (VC)}~\cite{w3c_vc} is a data structure that represents cryptographically verifiable and tamper-proof claims about an entity. A claim is a statement about an entity representing an attribute such as name, address, or birth date. 
VCs allow users to store several claims in an interoperable format, cryptographically signed, and fully controlled by the user. Having verifiable credentials, the user can submit claims to the service provider.

\textit{Verifiable Presentation (VP)}~\cite{w3c_vc} is a collection of one or more verifiable credentials from one or more issuers shared with the service provider. It is just a way to represent a group of different credentials chosen by the user rather than using several verifiable credentials independently. A digital signature can cryptographically verify the authorship of the collection of credentials.

Finally, a \textit{Wallet} is a software application that stores and manages identity data, VCs, and cryptographic keys. The key features include: user-friendly interface, secure storage of sensitive information, and integration interfaces for other applications and services.

\begin{figure}[ht]
\includegraphics[width=0.70\columnwidth]{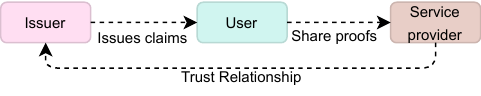}
\centering
\caption{Self-sovereign identity roles.}
\label{fig:ssi-roles}
\end{figure}

Figure~\ref{fig:ssi-roles} illustrates the general architecture of an SSI management system: Issuer, User, and Service Provider.

The \textit{Issuer} creates and issues VCs, verifies identity claims, generates VCs with appropriate signatures, and manages issuer credentials.
The \textit{User (Identity Holder)} owns and controls their identity data, generates and controls the cryptographic keys, issues verifiable credentials, and presents credentials to relying parties.
The \textit{Service Provider (Verifier)} receives and validates VCs, verifying their authenticity and integrity. It evaluates VC content against specified criteria and makes decisions based on verified information.

The concepts associated with Zero-Knowledge Proofs (ZKPs) are important for understanding the properties we can achieve in our system by using ZKP Verifiable Credentials. The ``Compound Proof" is a mechanism that is not currently implemented in our system prototype, but it is something to consider for future implementation. ``Selective Disclosure" is what allows our system to create Verifiable Presentations (VPs) using the Verifiable Credentials (VCs) already present in the user's wallet. The remaining concepts are properties utilized in our system due to the use of ZKP credentials, and it might be reasonable to omit their descriptions. However, in my opinion, it is important to include the descriptions to understand the capabilities that ZKP provides.

The use of VCs does not necessarily expose user data required for verification.
Zero-Knowledge Proofs (ZKPs)~\cite{zkp} can provide privacy by verifying information without revealing the underlying data.
This is achieved through cryptographic techniques that prove a statement true without disclosing specific details. 
For example, ``Selective Disclosure" allows the creation of VP using the VC in the user's wallet, revealing only necessary information while keeping other data private.

\subsection{Blockchain and Identity}

Blockchain technology offers key properties for building robust solutions~\cite{beginning_blockchain_2018}:
immutability, double-spend resistance, consistency, transparency, decentralized governance, and resilience.

\textit{Immutability} ensures that once a transaction is recorded, it is highly resistant to alteration. The integrity can be compromised by a 51\% attack~\cite{aponte202151}, where an entity with majority control of the computing power can perform alterations.
\textit{Double-spend resistance} prevents the same amount from being spent multiple times by ensuring prior reception of the intended payment amount.
Ledger \textit{consistency} is achieved through consensus mechanisms, maintains a stable decentralized system even with some fraudulent transactions and node failures.
\textit{Transparency} is facilitated by the sequential chain structure, linking each block back to the genesis block for comprehensive record-keeping, allowing the system to be audited whenever necessary.
The \textit{decentralized governance} eliminates central authority, empowering participants with decentralized decision-making, akin to a form of ``democracy''.
Finally, \textit{resilience} is ensured by the distributed and decentralized architecture, which withstands node failures, network latency, and other faults.

There has been considerable research work on using blockchains to support decentralized authentication and authorization of digital identities. 
Notable systems include: Sovrin, uPort, ShoCard, and Civic.

Sovrin~\cite{sovrin_technical} is a public, permissioned identity network built on Hyperledger Indy that supports privacy-preserving, verifiable credentials to be used in digital interactions.
Its trust relies on the reputations and non-collusion of its nodes.
The ledger is accessible to the public and is operated by trusted stewards vetted by a non-profit foundation, coordinating support from reputable companies, non-profits, and academic institutions.
Sovrin can utilize zero-knowledge proof (ZKP) techniques for enhanced privacy but it lacks a decentralized storage system for supporting document-based validation processes.

uPort~\cite{uport_whitepaper} is an identity management system based on Ethereum and IPFS. uPort identities are implemented by smart contracts deployed on Ethereum while the credentials and profile data are stored on the IPFS. A Registry Contract offers a cryptographic link between an on-chain identifier and its corresponding off-chain data. Furthermore, uPort provides wallets for users to facilitate the management of their identities.

ShoCard~\cite{first_look_comparison} is a platform that enables the creation of a digital identity card on a mobile phone bound to identity attributes. It uses Bitcoin as a timestamping service to store user identities and logs. However, ShoCard does not adhere to W3C standards or utilize ZKP mechanisms.

Civic~\cite{civic_whitepaper} operates within a blockchain-based identity authentication system, where a third party wallet generates key pairs and stores user identification data locally. It leverages the Ethereum blockchain and uses smart contracts to monitor the proof of attestation. Civic incorporates W3C standards and ZKP mechanisms but does not use a permissioned blockchain, identity document storage systems, or audit logs features.

\subsection{Decentralized Storage}
\label{sec:decentralized-storage}

Blockchains are inefficient for storing large data volumes due to scalability, privacy, and storage limitations~\cite{yang2020review}. Their immutability worsens these issues, as data cannot be deleted once stored, creating challenges in data protection and management.

Decentralized storage systems, such as IPFS~\cite{benet_ipfs_2014}, Filecoin~\cite{filecoin}, and Storj~\cite{storj_whitepaper},  distribute data across a network of nodes, rather than relying on centralized servers. This architecture inherently enhances data security by reducing the risk of data breaches and censorship. Additionally, it provides a high degree of availability and fault tolerance, as data is replicated across multiple nodes.

Data is fragmented into smaller pieces and distributes them across multiple nodes~\cite{data_fragmentation}.  Combined with robust encryption, this method enhances data security and makes it significantly more difficult for unauthorized parties to access or recover information.

However, decentralized storage also presents challenges. Data retrieval can be more complex and time-consuming compared to centralized storage. Additionally, ensuring data integrity and availability requires careful system design and management.

Decentralized storage systems while offering advantages in terms of data availability and fault tolerance, face their own hurdles when it comes to data deletion\cite{secure_delete}. The distributed nature of these systems makes it impractical to guarantee complete data eradication due to the potential for data replication across multiple nodes.

\textit{Storj}~\cite{de2021exploring} is a decentralized storage platform that divides files into encrypted shards distributed across a network of nodes, ensuring no single entity possesses a complete copy of the data. This enhances security and data management. 
Storj consists of uplinks, storage nodes, and satellites. Uplinks are client-side applications for uploading, downloading, and managing files, handling encryption and data segmentation. Storage nodes, operated by individuals providing excess hard drive space, store and retrieve data, earning tokens. Satellites manage metadata, audit integrity, repair data, and handle payments, ensuring network reliability.
For data deletion, the Uplink validates the user's permission with the Satellite, which generates signed agreements for each segment's deletion. Storage nodes confirm deletion through signed messages. The Satellite updates metadata, stops charging the customer, and ceases payments to nodes. Periodic garbage collection ensures any missed deletions are completed, maintaining storage efficiency and accuracy.

The Storj has mechanisms for ensuring the delete operation. 
When a user initiates a delete request, the Uplink first validates the user’s permission with the Satellite. The Satellite then generates signed agreements for the deletion of each segment, ensuring storage nodes acknowledge and proceed with the deletion request. Storage nodes confirm the deletion by returning signed messages, which indicate that the files and their associated bookkeeping information have been removed or were previously deleted. The Satellite updates its metadata by removing the segment pointers, ceasing to charge the customer, and stopping payments to the storage nodes. Additionally, Storj employs periodic garbage collection to clean up data from nodes that may have missed initial delete messages, thereby maintaining storage efficiency and accuracy.

\section{SoverClaim Overview}
\label{sec:soverclaim-overview}

Let us describe a document-based claim process supported by SoverClaim. As an example, we will focus on a health insurance use case.
Decentralized applications have the potential to give individuals greater control over their health data, enhancing security, privacy, and interoperability~\cite{Houtan2020ASO}. 
By decentralizing health data storage, the risk of data breaches is reduced, and secure data exchange between patients, insurers, and healthcare providers is facilitated.

In this scenario, the patient is the \textit{User}, the insurance company is the \textit{Issuer}, and a medical hospital is the \textit{Service Provider}.
The following steps illustrate the process:

\begin{enumerate}[label=\roman*.]
    \item \textbf{Patient Secure Connection}: The patient establishes a secure connection with the insurer using SoverClaim;
    \item \textbf{Document Upload}: The patient uploads the required documents (e.g., images) to the decentralized storage system of SoverClaim;
    \item \textbf{Retrieving Document URL}: After the documents are uploaded, the patient obtains a URL for each uploaded document;
    \item \textbf{Issuance of DID}: The insurer issues a Decentralized Identifier (DID) to the patient, which serves as a unique identifier for the insurance credentials;
    \item \textbf{Verifying Insurance Credential}: The patient requests the insurer to validate their uploaded insurance documents. The insurer verifies the documents and issues a verifiable insurance credential to the patient;
    \item \textbf{Credential Presentation}: When the patient visits a medical service provider, they present the credential;
    \item \textbf{Provider Verification}: The medical service provider verifies the patient's insurance credential using SoverClaim, ensuring the credential's authenticity and validity;
    \item \textbf{Secure Access and Deletion}: The decentralized storage system ensures that all documents are securely stored and can be accessed.
    After they are no longer needed, they are deleted to comply with data protection laws.
\end{enumerate}

\begin{figure}[ht]
\centering
\includegraphics[width=0.85\columnwidth]{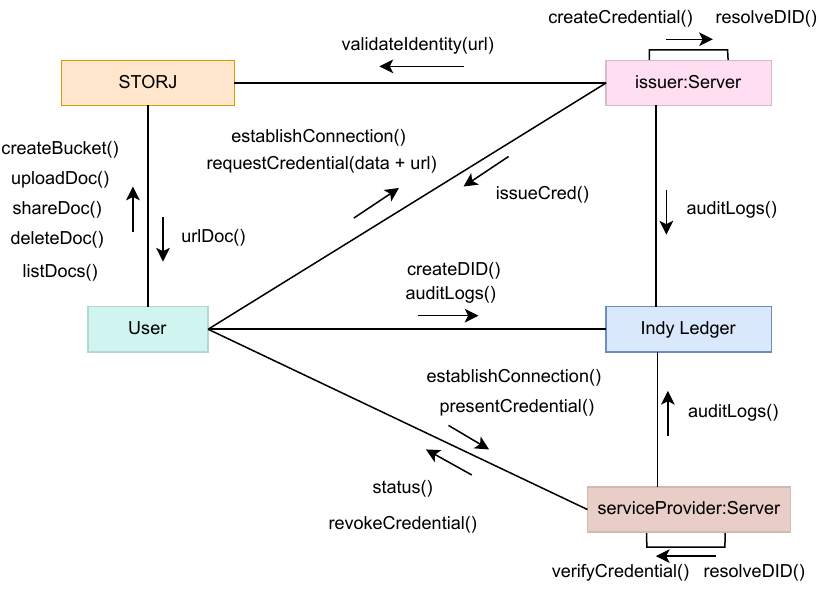}
\caption{SoverClaim in action.} 
\label{fig:methods}
\end{figure}

Figure~\ref{fig:methods} provides a collaboration diagram of SoverClaim in action along with their corresponding operation names.

\section{SoverClaim Implementation}
\label{sec:impl}

Given the illustrative use case, SoverClaim needs to provide the following features:
\begin{itemize}
    \item DID Creation, allowing the User to create decentralized identifiers, exchange nonpublic DIDs, and establish a secure peer-to-peer connection using DIDComm protocol;
    \item Identity document management, allowing the User to upload, store, delete, and share their documents on the storage network;
    \item Credential issuance, the User can request the Issuer for a credential, with its data and the correspondent document if needed, and the Issuer can validate the provided information with the shared document;
    \item Credential presentation, using a presentation proof with the information from its credentials;
    \item Audit logs, register when the User requests a credential, or sends a presentation;
    \item Credential revocation, executed when credentials expired or when the User requests it.
\end{itemize}

\begin{figure}[ht]
\centering
\includegraphics[width=0.75\columnwidth]{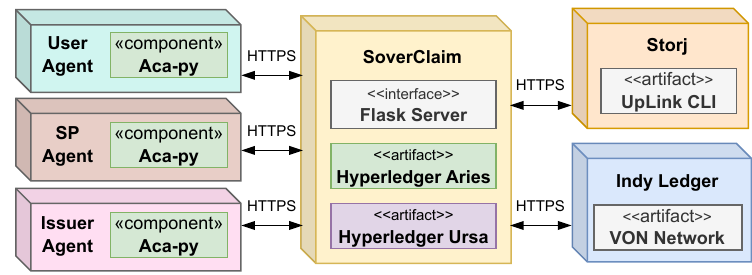}
\caption{SoverClaim deployment diagram.} 
\label{tech1}
\end{figure}

The architecture of SoverClaim incorporates several components and technologies, represented in the deployment diagram shown in Figure~\ref{tech1}. 

\subsection{Self-Sovereign Identity Blockchain}

The SSI blockchain chosen was Indy Ledger~\cite{hyperledger_indy} due to its ability to create and use decentralized identities, compatible with W3C DID and VC standards.
Indy operates as a permissioned blockchain and can securely record and validate transactions related to decentralized identities and provides transparent audit logs.

There are several Hyperledger libraries involved: \textit{Indy} is the ledger, \textit{Aries} provides communication functions, and \textit{Ursa} is the cryptographic library used for the credentials.

VON (Verifier's Open Network)~\cite{vonnetwork} is an Indy ledger testnet.
It is part of the Indy project and serves as a sandbox environment for developing and testing decentralized identity solutions using the Indy framework.
VON can be set up a private ledger for experiments.
Each node on the network runs in a separate Docker container that simplifies deployment and  deployment. 
Additionally, there is also the Ledger Browser, a tool that allows the monitoring of network nodes and provides the ability to browse, search, and filter ledger transactions.

\subsection{Decentralized Document Storage}

Storj~\cite{storj_whitepaper} was chosen for SoverClaim due to its decentralized architecture, end-to-end encryption, and data fragmentation mechanism.
The data is distributed across a vast network of nodes and this aligns well with the decentralization application principles.
Storj also has good tools and documentation and supports document deletion.
With appropriate metadata management, users can exercise granular control over the lifespan of their claim documents.

\subsection{Orchestration}

The SoverClaim prototype, developed using Python 3~\cite{python}, 
incorporates controllers for Aries agents, which serve as autonomous entities or software components that represent individuals, organizations, or assets within the decentralized identity context. A fundamental component of this implementation is Aca-py~\cite{aca-py}, also known as ``Aries Cloud Agent for Python'' which facilitates secure communication, credential issuance, and verification in SSI systems.

The prototype uses the requests library for sending and receiving HTTPS requests and working with the API of Aca-py. Additionally, Flask~\cite{flask}, a popular Python web framework, is used to set up web servers that handle incoming webhook events, including message reception and credential requests from Aries agents.

\subsection{DID Protocol}

In SSI, entities can generate both public and non-public DIDs to accommodate the representation format of a verifiable credential. 
Public DIDs use the \texttt{did:sov}~\cite{did_sov} operation, which requires a ledger, in this case, the Indy ledger, to resolve the DID. 
This involves querying the ledger to retrieve the corresponding DID document to obtain its public key. 
Non-public DIDs utilize the \texttt{did:key}~\cite{did_key} operation, which does not depend on a distributed ledger for resolution. 
The self-contained DID document, containing the public key, resolves locally. 

\textbf{DIDComm}~\cite{did_comm_git}, short for ``Decentralized Identifiers Communication'', is a protocol specification designed to enable secure and private peer-to-peer communication between entities using decentralized identifiers.
SoverClaim uses DIDComm to exchange credentials, presentations, non-public DIDs, authentication or authorization status, and messages.
Establishing a DIDComm connection between the user and issuer follows these steps: 
a) The issuer creates an out-of-band invitation sent to the user using QR codes or URLs; 
b) The user receives the invitation and sends a connection request to the issuer since both parties should agree on creating the connection; 
c) The issuer then receives the request, accepts it, and sends a response to the user that establishes the peer-to-peer connection. An identical process occurs between the user and the service provider.

\subsection{Document Storage with Storj Protocol}

SoverClaim integrates UpLink CLI~\cite{storj_uplink} to ``talk'' with the Storj network.
With it, users can create file storage buckets, upload files, list files within a bucket, share and delete specific files.

To interact with these operations within the SoverClaim application, users need to provide the name of the bucket and the claim document's file path. Then, they can perform the desired actions, such as uploading, listing, deleting, or sharing the claim documents. When sharing files, users will receive a URL to access the document.

To \textit{upload} a file into Storj, a bucket is created with the command uplink mb sj://$<$bucket name$>$. Then, the file present on your own device is uploaded by providing the bucket name and the file path using the command uplink cp $<$file path$>$ sj://$<$bucket name$>$. 

To \textit{list} all files inside the bucket, the following command is used: uplink ls sj://bucket name.

To \textit{share} a file, the user needs to create a URL for sharing the uploaded file by using the
command uplink share --url --not-after=none sj://bucket name/file name. This URL gives read permissions to any User that accesses it, unless extra flags are used in the previous command. The flag “–url” allows Storj to generate the URL, and “–not-after=$<$time$>$” can be used for disallow access after the specified time. 

To \textit{delete} a file from Storj, the User must request the deletion of the file using the command uplink rm sj://bucket name/file name. The command asks for all the segment pointers related to the file to help locate where the fragments are stored. Then, each storage node replies with a signed message, confirming either that they received the delete request and will delete the file or that it was already deleted. Lastly, the segment pointers are removed.

\subsection{Issue Credential Protocol}

\begin{figure}[ht]
\centering
\includegraphics[width=0.87\columnwidth]{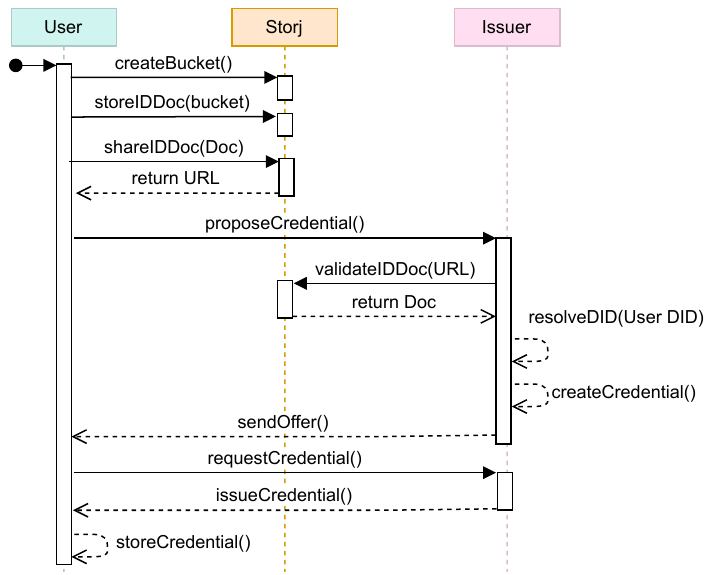}
\caption{Issue credential protocol} 
\label{seq_issue}
\end{figure}

The credential protocol used by SoverClaim uses Aries RFC 0454: Present Proof Protocol 2.0~\cite{present_proof_v2}. 
This protocol standardizes the process for a service provider to request and receive a verifiable presentation from a user. When the user establishes a secure connection with the service provider, the credential protocol starts as shown in Figure~\ref{seq_issue}.
The protocol works as follows; It starts with the user interacting with Storj to upload the desired claim document, which will be required to validate the credential data. Once the file is successfully uploaded, the user receives a shared URL that points to the file in Storj. Having exchanged DIDs and obtained the URL, the user sends a proposal to the issuer, including the desired claims for the credential and the shared URL of the identity document. The issuer then proceeds to validate the authenticity of the identity document to verify the data provided by the user. Additionally, they resolve the user's DID to confirm ownership of the provided non-public DID. Upon successful validation, the issuer sends a credential offer to the user, describing the details of the credential they intend to provide. The user confirms the offer and requests the issuance of the credential. The issuer is now prepared to generate the credential and send it to the user, completing the process.

\subsection{Present Credential Protocol.}

\begin{figure}[ht]
\centering
\includegraphics[width=0.60\columnwidth]{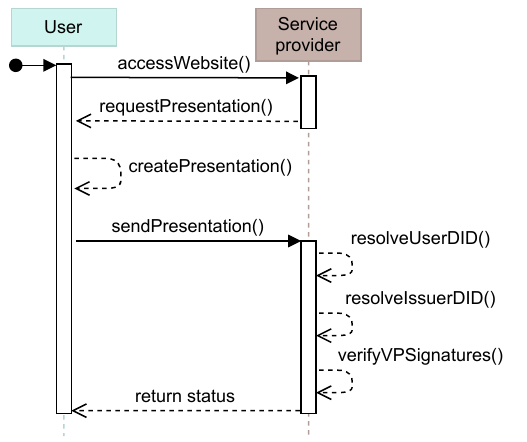}
\caption{Present Credential protocol.} 
\label{seq_presentation}
\end{figure}

The credential protocol used by SoverClaim is based on Aries RFC 0454: Present Proof Protocol 2.0~\cite{present_proof_v2}. 
This protocol establishes a standardized method for a service provider to request a presentation from a prover, and for the prover to respond by presenting a proof to the verifier.
After the user establishes a secure connection with the service provider, the Present Credential protocol begins as represented in Figure~\ref{seq_presentation}.
The protocol starts when a user accesses a service from the service provider's website, which replies by requesting a presentation proof from the user. After receiving the request, the user reviews all their available credentials and chooses one or more that satisfies the specific claims requested by the service provider.
A ZKP mechanism can guarantee that no superfluous information is disclosed during this selection process.
Then, the service provider receives the verifiable presentation and initiates the verification process. This process involves resolving the DIDs of both the user and the issuer to confirm their ownership and to validate the signatures of the verifiable credential.
The verification result is then provided to the user, determining their access to the requested resources.

\subsection{Audit Logs Protocol}

The audit log mechanism captures and records specific events, such as requesting a credential, receiving a new credential, sending a verifiable presentation, and receiving status updates from the service provider.
Each log entry, represented as a JSON data structure, contains the following attributes: operation type, credential type (e.g., Permanent Residence, Insurance Policy), user's DID, service provider or issuer's DID, and timestamp. This log allows users to have a proof of their online activity while enhancing trust in their identity to the auditors.

SoverClaim implements a process in which logs are encrypted and decrypted with a symmetric key, known as a session key, before being added to the blockchain.
To ensure confidentiality and user privacy, the session key is encrypted with the user's public key before being included alongside the encrypted logs in the blockchain.
This approach enhances security, allowing only the user with the corresponding private key to decrypt and access the logs.

When an auditor wants to view log content, they must request decryption information from the user. 
The user retains control over their log data and can choose whether to share it.
If so, the user securely transmits the relevant log details back to the auditor.
SoverClaim acts as a trusted intermediary for encryption and decryption while ensuring that the user's private key remains under their exclusive control.
The private key is stored locally using key stores, which encrypt and protect it using a passphrase.
The system shares the users' public keys with relevant entities as necessary, and it generates the session key locally and stores it alongside the audit log.

\section{Evaluation}
\label{sec:eval}

For the evaluation, we set up a cloud infrastructure to conduct experiments aimed at answering the following questions:
\begin{itemize}
    \item \textbf{Q1:} How does SoverClaim perform when each component and entity of the system is geographically distributed and used by multiple users? 
    \item \textbf{Q2:} What are the storage, CPU, and memory requirements to run SoverClaim?
    \item \textbf{Q3:} What is the monetary cost of running SoverClaim? 
\end{itemize}

Next, we describe the setup and present the experiment results.

\subsection{Setup} 

To evaluate the performance of our prototype, the system was deployed using Google Cloud instances, closely resembling a real-world scenario.
The setup consisted of five Linux virtual machines, each serving distinct roles: User, Issuer, Service Provider, SoverClaim Server, and VON Network (Indy ledger).
Each virtual machine runs Debian 11, featuring two virtual CPUs, 8 GB of RAM, and standard persistent disks (i.e., HDDs) for the boot disk.
These resources were provisioned in the Compute Engine environment from Google Cloud.
We deployed the virtual machines in different regions of Europe (London, Frankfurt, Netherlands, and Belgium) to ensure a geographically relevant setup for evaluating our prototype's performance and behavior.
Additionally, we used Locust~\cite{locust}, a load-testing tool that simulates multiple user requests and provides performance measures. 

\subsection{Performance} 

\begin{figure}[!ht]
  \centering
  \includegraphics[width=0.70\columnwidth]{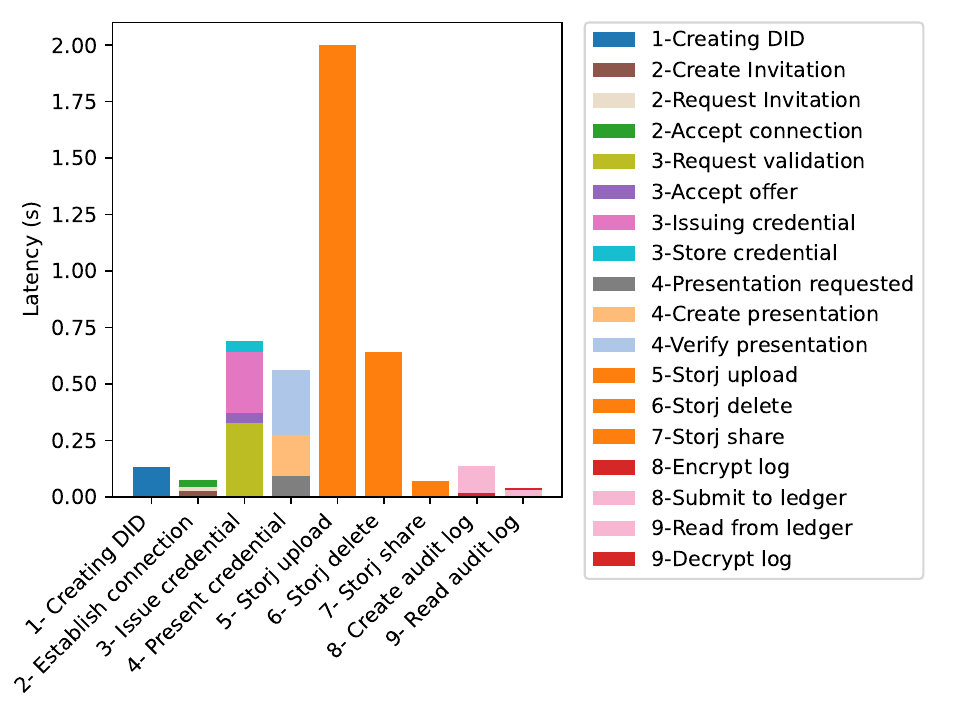}
  \caption{Latency evaluation of different functionalities.}
  \label{fig:portions}
\end{figure}

Our focus was to assess the latency of the interactions.
Figure~\ref{fig:portions} shows the results of the various interactions: Creating DID; Establishing a connection; Storj interactions; Issuing and Presenting a credential; Create and Read an audit log. Additionally, the breakdown of each interaction represents all individual processes.

\subsubsection{Resource Usage}

We used the monitoring capabilities of the virtual machine instances provided by the Google Cloud Computer Engine, namely, Storage, CPU, Memory Requirements.

During our analysis, we observed the issuer agent's behavior while handling a workload of credential issuance requests from multiple users.

\begin{figure}[!ht]
  \centering
  \includegraphics[width=0.70\columnwidth]{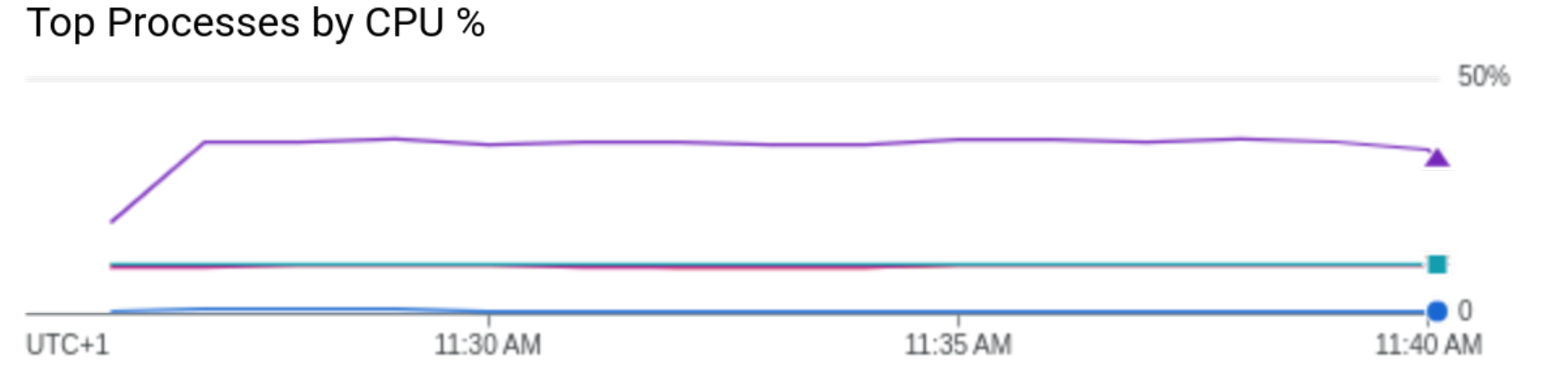}
  \caption{Processes by CPU percentage.}
  \label{fig:cpu}
\end{figure}

In Figure~\ref{fig:cpu}, we can analyze the following percentage of CPU usage for each process. 
Each node in our blockchain infrastructure represents 10\% of CPU usage, resulting in a cumulative total of 40\% since we operated four nodes within the network (represented in green). 
The Aries agent, which played an integral role in our system, accounted for 36\% of CPU usage (represented in purple). 
Additionally, the Secure Shell Daemon (sshd), which facilitates secure remote access over the network, consumed 1\% of CPU usage (represented in blue). 

\begin{figure}[!ht]
  \centering
  \includegraphics[width=0.70\columnwidth]{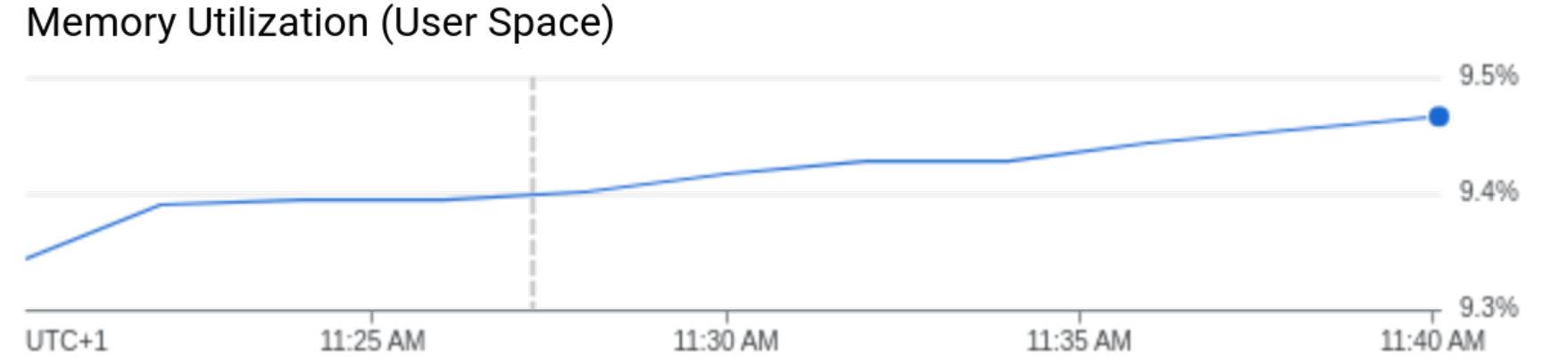}
  \caption{Memory used by Issuer agent.}
  \label{fig:memory}
\end{figure}

In Figure~\ref{fig:memory}, we can see the percentage of total memory (8~GB) used by the issuer when handling the request for issue credentials.
Additionally, we monitored the storage consumption of the agent during approximately 3500~requests of workload.
As a result, around 56~MB of data was generated, which includes records, logs, and stored events. This accounts for around 16~KB per request.

\subsection{Operation Cost}

Our estimates are based on Google Cloud's monthly cost, considering the creation and utilization of a virtual machine with the specified characteristics mentioned earlier.
Each machine costs approximately 50~€ per month (around 0.06~€ per hour).
For our prototype, the expenses were: using one machine to run SoverClaim, one for the private blockchain, and three machines, one for each entity of the SSI system, making a total of 5~machines.
As a result, the overall monthly cost of operating our prototype would amount to 250~€, with an additional 50~€ per entity.
In a real-world scenario, each entity would operate its device to run SoverClaim, and the responsible nodes would ensure the blockchain's functionality through decentralization. Consequently, this would significantly reduce the cost of running our SoverClaim prototype to just 50~€ per month.

\subsection{Discussion}

Regarding Q1, SoverClaim performs efficiently even when its components and entities are geographically distributed and used by multiple users.
The latency for creating DIDs, establishing connections, using Storj, issuing and presenting credentials, and managing audit logs is low enough for interactive user interfaces.

Regarding Q2, the resource requirements are manageable within the capabilities of mid-range cloud servers.

Regarding Q3, the monetary cost of running on the cloud is reasonable.
Given that claim processes occur occasionally, the costs can be shared between multiple users, making it even more economically feasible.

\section{Conclusion}
\label{sec:conclusion}

This work has explored the potential of decentralized applications to enhance privacy, security, and user control in the management of digital identities and sensitive documents.

Our prototype application, SoverClaim, integrates the blockchain and decentralized storage technologies to handle identity management and document submissions, with transparency while adopting standard W3C protocols.

In our experiments, SoverClaim was able to issue and present credentials in less than 750~ms and 600~ms, respectively. These times demonstrate that the application is efficient and can be made practical.

Future work will focus on extending SoverClaim to handle more use cases and conducting more extensive evaluations in real-world applications, and showing how we can move towards a digital future where individuals have greater control over their personal data and can interact with online services securely and privately.

\section*{Acknowledgments}

This work was partially supported by national funds through Funda\c{c}\~ao para a Ci\^encia e Tecnologia (FCT) with reference UIDB/50021/2020 (INESC-ID), by the European Union’s Horizon 2020 research and innovation programme under grant agreement No 952226, project BIG (Enhancing the research and innovation potential of Tecnico through Blockchain technologies and design Innovation for social Good), and by Project Blockchain.PT – Decentralize Portugal with Blockchain Agenda, (Project no 51), WP 1: Agriculture and Agri-food, Call no 02/C05-i01.01/2022, funded by the Portuguese Recovery and Resilience Program (PRR), The Portuguese Republic and The European Union (EU) under the framework of Next Generation EU Program.

\bibliographystyle{unsrt}  
\bibliography{paper}

\end{document}